\documentstyle[12pt]{article}
\begin{document}
\baselineskip=21.5pt

\begin{center}

{\Large{Blume-Emery-Griffiths Model on the Square Lattice with 
Repulsive Biquadratic Coupling}}
\footnote{Work partially supported by the Brazilian Agencies CNPq, 
FINEP and CAPES}

\vskip2cm

{\large{N. S. Branco\footnote{e-mail: nsbranco@fsc.ufsc.br}}} \\
     Universidade Federal de Santa Catarina, Depto.\ de F\'{\i}sica,
      88040-900, Florian\'{o}polis, SC, Brazil.
\end{center}

\vskip1.5cm

\begin{abstract}

   Using a real-space renormalization group procedure with no 
adjustable
parameters, we investigate the Blume-Emery-Griffiths model on 
the square lattice. The formalism respects sublattice symmetry,
allowing the study of both  signs of $K$, the biquadratic exchange
coupling. Our results for $K>0$ are compared with other renormalization
group calculations and with exact results, in order to assess the 
magnitude of the errors introduced by our approximate calculation. The
quantitative agreement is excellent; values for critical parameters
differ, in some cases, by less than $1 \%$ from exact ones. For $K<0$,
our results lead to a rich phase diagram, with antiquadrupolar and
ferromagnetic ordered phases. Contrarily to Monte Carlo simulations,
these two phases meet only at zero temperature. Both 
antiquadrupolar-disordered and ferromagnetic-disordered transitions 
are found to be continuous and no ferrimagnetic phase is found.  

\vskip 1.5cm


\vskip 0.7cm

 
\end{abstract}
\newpage

\baselineskip=21.5pt
\parskip=3pt

\section{Introduction}
                  
     The most general spin-1 Ising model with up-down symmetry is 
the Blume-Emery-Griffiths (BEG) model, which Hamiltonian reads:
\begin{equation}
   {\cal  H} = - J \sum_{<i,j>} S_i S_j - K \sum_{<i,j>} S_i^2 S_j^2
   + \Delta \sum_i  S_i^2, \;\;  
   \label{eq:hamil} 
\end{equation}
where the first two sums are over all nearest-neighbor 
pairs on a lattice, the last one is over all sites 
and $S_i = \pm 1,0$ \cite{BEG,ber1}. This model 
was originally introduced to study phase separation and superfluidity
in $^3$He-$^4$He mixtures \cite{BEG}. Later it has been applied to 
describe 
properties of  multicomponent fluids \cite{exp1}, microemulsions 
\cite{exp2}, semiconductor alloys \cite{exp3}, and electronic 
conduction
models \cite{exp4}.  

   From the theoretical point of view, the BEG model was 
extensively studied for positive biquadratic exchange ($K>0$). Its
phase diagram is well understood: it exhibits two disordered phases
and a ferromagnetic phase. The transition between them
can be a continuous or a first order one, with the presence of 
an ordinary tricritical line, an isolated critical line, and a
line of critical end points \cite{ber1}. 

    Recently, attention has been drawn to the repulsive
biquadratic ($K<0$) model. In this case, sublattice symmetry
may be broken and the possible phases are classified according
to the values of $M_A \equiv \langle S_i \rangle_A, \; 
M_B \equiv \langle S_i \rangle_B, \; Q_A \equiv \langle S_i^2 \rangle_A,$
and $Q_B \equiv \langle S_i^2 \rangle_B,$ where the subscripts
$A$ and $B$ refer to the two sublattices of a bipartite lattice
(which we limit ourselves in this work).
A mean-field solution, firstly proposed
by M. Tanaka and T. Kawabe \cite{japa} and later extended 
by W. Hoston and A. N. Berker \cite{ber2}, leads to a rich phase diagram,
with two new ordered phases: antiquadrupolar ($M_A=M_B=0, \; Q_A \neq Q_B$) 
and  ferrimagnetic ($0 \neq M_A \neq M_B \neq 0$ and
$Q_A \neq Q_B$). Moreover, the $(J, \; K, \; \Delta)$ phase diagram 
shows a variety of multicritical
points, such as critical end points, bicritical, tetracritical, etc.
\cite{japa,ber2}. However, results obtained using
mean-field approaches are expected
to hold only at high dimensions; in what concerns
the BEG model, discrepancies
have been found between results obtained by the cluster-variational
method \cite{anders1} and by Monte Carlo  simulation \cite{anders2}:
both predict more than one ferrimagnetic phase for the
three-dimensional model while the mean-field approach \cite{ber2}
detects just one ferrimagnetic phase. In two dimensions, previous 
studies \cite{japa,wang,ber3} found no 
ferrimagnetic phase, although they disagree in what concerns the
shape of the boundary between the ferromagnetic and the antiquadrupolar
phases. While in  Ref. \cite{japa} it is found that there is
a first order transition between these two phases, the results of Refs.
\cite{wang,ber3}
suggest that the phases meet only at zero temperature and that
the antiquadrupolar-disordered (AD) and ferromagnetic-disordered (FD)
transitions are continuous.

    Since the question is not yet settled, an independent
calculation is desirable. In order to address these questions, we 
apply a real-space renormalization group (RG) procedure to
study the BEG model on the square lattice. Our formalism allows the
study of both attractive ($K>0$) and repulsive ($K<0$) biquadratic
interaction. Although the former is now well understood, we
also present our results for this case: comparing them 
with more precise or with exact results, we can infer the errors
introduced by the approximate procedure we use.

   Before we carry on with a brief explanation of the formalism, let
us review some of the exact information and symmetries 
associated with the model:

\begin{enumerate}

\item[(i)] on bipartite lattices, the antiferromagnetic model ($J<0$) 
can be mapped onto the ferromagnetic one ($J>0$) by flipping 
every spin on one sub-lattice. So, we can restrict ourselves
to the ferromagnetic case without loss of generality;

\item[(ii)] $\Delta \ll -1$: in this region, those configurations 
with $S_i = \pm 1$ dominate  the ensemble averages and the spin-$1/2$
Ising model is reobtained;

\item[(iii)] $\Delta \gg 1$, $J$ and $K$ finite: all spins are in the
zero state and both $M$  and $Q$
are zero;

\item[(iv)] $\Delta \gg 1$ and  $2(J+K) \sim \Delta$; either the 
configuration with all spins $S_i=0$ or the configuration
with all spins $S_i=+1$ (or, equivalently, $S_i=-1$) is the
ground state. Their energies are $E/N (\{S_i=0\}) = 0$ and
$E/N (\{S_i= \pm 1\}) = 2(J+K)-\Delta$; so, there is a
first order transition at $2(J+K) = \Delta \gg 1$; 

\item[(v)] {\it Griffiths symmetry}: at $J=0$ we can define
a new variable for each site $i$, $t_i = 2 S_i^2-1$. The
Hamiltonian is then transformed into (apart from additive
constant terms):
\[ {\cal H} = - J_G \sum_{<i,j>} t_i t_j +  H_G \sum_i t_i,
\;\; t_i = \pm 1, \]
with $J_G = K/4$ and $H_G = K + \frac{1}{2}(\ln 2 - \Delta)$
(on the square lattice). For $K>0$ ($K<0$),
this is the ferromagnetic (antiferromagnetic) spin-$1/2$ Ising 
model on a uniform magnetic field. Thus, in an  exact renormalization 
group procedure, one expects to obtain 
$(J=0,\; K=1.763,\; \Delta=4.219)$ and \mbox{$(J=0,\; K=-1.763,\; 
\Delta=-2.832)$} as critical fixed points for the ferromagnetic and 
antiferromagnetic model on the square lattice, respectively.

\item[(vi)] {\it Three-state Potts model}: for $K = 3J$ and
$\Delta = 8J$, the BEG Hamiltonian reduces to the ferromagnetic 
three-state Potts model \cite{ber1}. On the other hand, 
\mbox{$K = -3J$} and 
$\Delta = -8J$ is equivalent to the antiferromagnetic 
three-state Potts model \cite{ber2}. The exact location
of the ferromagnetic model fixed point would be
\mbox{$(J=0.5025,\; K=1.5076,\; \Delta=4.0202)$}, while the 
corresponding
fixed point for the antiferromagnetic model is at 
\mbox{$(J=\infty,\; K=\infty,\; \Delta=\infty)$} \cite{Wu}.

\end{enumerate}

    We expect a reliable RG procedure to respect, at least in
an approximate way, these symmetries, {\it which are not}
a priori {\it incorporated in the formalism we apply in this work}.

   The remaining of the paper is organized as follows. In Section 2
we explain the RG formalism, in Section 3 we present the results, 
for both  $K<0$ and $K>0$, and finally in Section 4 we briefly review
the results obtained in  this article.

\section{Formalism}

    As is usual in small-cell RG approaches, we approximate the Bravais
lattice (in our case, the square lattice) by an appropriate
hierarchical lattice. As long as the symmetries of the ground-states
are preserved, no spurious results  are introduced. In the present work,
this requires a cell which respects sub-lattice symmetry, in order 
to get the correct behavior for $K<0$. We note that the results
obtained are {\it exact} on the chosen hierarchical lattice but only
approximate on the Bravais lattice; in particular, one does not
expect to obtain results as precise as  those achieved using
Monte Carlo simulations or conformal invariance arguments 
(although the precision obtained in the present work
is excellent). More generally, it is known that two-dimensional
real-space RG procedures yield more accurate results than their
three-dimensional counterpart (see, for instance, \cite{rsrg} and
references therein).
We are here mainly interested in qualitative
features of the phase diagram, like, for instance, the presence 
of distinct phases and universality classes. 

     The cell chosen (see Figure 1) has been used with success 
in many studies of antiferromagnetic systems (\cite{muri,dico} and
references therein). We then impose that the correlation function 
between the two terminal sites of the original and
renormalized graphs are preserved \cite{tsallis}:
\begin{equation}
  \exp (- \beta {\cal H}_{12}) = Tr \; 
\exp (- \beta {\cal H}_{123456}), \label{eq:rg} 
\end{equation} 
where $Tr$ means a partial trace over the internal sites of the
cell (those marked with 3, 4, 5 and 6 in Figure 1).
We rewrite the Hamiltonian as a sum of ``bond'' terms
(from now on, the factor  $-\beta$ will be absorbed into the 
interaction parameters):
\begin{equation}
   {\cal H}_{12} = - J' S_1 S_2 - K' S_1^2 S_2^2 + 
\frac{\Delta'}{4} (S_1^2+S_2^2) + G', \label{eq:cell2}
\end{equation}
and:
\begin{eqnarray}
{\cal H}_{123456} & = - J (2 S_1 S_3 + S_1 S_4 + S_3 S_6 + S_3 S_5 +
 S_4 S_6 + S_2 S_5 + 2 S_2 S_6) \nonumber \\ 
& - K (2 S_1^2 S_3^2 + S_1^2 S_4^2 + 
S_3^2 S_6^2 + S_3^2 S_5^2 + S_4^2 S_6^2 + S_2^2 S_5^2 + 2 S_2^2 S_6^2)
\nonumber \\
& + \frac{\Delta}{4} ( 3 S_1^2 + 4 S_3^2 + 2 S_4^2 + 2 S_5^2 + 4 S_6^2 
+ 3 S_2^2) , \label{eq:cell1}
\end{eqnarray} 
where primed quantities are renormalized parameters and $G'$ is
necessary to  redefine the zero of energy.

  Note that this way to write the cell  Hamiltonian is equivalent
to  attribute weights to the sites in the one-site (crystal-field
$\Delta$) interaction, according to their coordination number.
This is necessary for finite lattices to approximate correctly the
infinite lattice behavior (see, for instance, \cite{dico}).

   In this way, we obtain the RG relations between original and 
renormalized parameters:
\begin{equation}
J' = J'(J,\; K,\; \Delta); \: K' = K'(J,\; K,\; \Delta); \: 
\Delta' = \Delta'(J,\; K,\; \Delta) . \label{eq:rgt}
\end{equation}
    The critical points are then evaluated as non-trivial fixed
points of the above relations and critical indices are obtained
linearizing the equations around these fixed points.

   As we shall see later, some of the symmetries of the model
are exactly reproduced by our procedure and some are reproduced to
a very good approximation.     

\section{Results}

\subsection{Positive K}

    For positive biquadratic interaction, the phase diagram is
qualitatively similar to the one found in \cite{ber1} (see  Figure 
2). One exception is the $OGF_2$ line, which is not a straight
line in our treatment; this means that the Griffiths symmetry is
not exactly respected (see item (i) in the first section). 
We show in Figure 3 the
exact curve (dotted line), slightly above the curve found in our work
(full line): note that the discrepancy is very small. It is easy to
show that the Griffiths symmetry is not respected due to surface
sites i.e., those sites at the boundary of the cell which have
less than 4 first neighbors. So, we expect that the
discrepancy will diminish as bigger cells are used. Nevertheless,
even with the small cells used, the agreement is very good, as
Figure 3 shows. In particular, the fixed point $G$ in  Table I
is within $2 \%$ of the exact value.
It is worthy to stress that, contrarily to some RG procedures,
we do not impose this symmetry {\it a priori}.

    Another model which is not exactly recovered in our formulation
is the three-state ferromagnetic Potts model. Nevertheless, the
corresponding fixed-point, $P$ in Table I, differs by less than 
$0.1 \%$ from the exact value. 

  In Table I we compare our 
evaluation of the fixed points with exact or previous results. 
As for $G$ and $P$, the agreement is a remarkable one, taking into
account the size of the cells used. Let us mention that our evaluation
of the critical exponents are not as precise as for the critical
points but are in  qualitative agreement with expected results. In
particular, we can correctly describe first-order phase transitions
associated with a discontinuity in $Q$, like the one cited at
$(iv)$ in the Introduction. In order to detect discontinuities
in $M$ a study of odd interactions is necessary; since we can determine 
the order of the transitions for $K < 0$ from general arguments, we 
limited ourselves to even interactions.

   Finally, let us mention that the symmetries $(i)$ to $(iv)$ in
the Introduction are exactly respected by our treatment.

\subsection{Negative K}

   For this case, competition between bilinear ($J$) and biquadratic ($K$)
couplings takes place and, as a result, a new phase is present in
two dimensions, namely the antiquadrupolar phase. Early works do not 
agree in what concerns the boundary (if any) between this phase and
the ferromagnetic one. Our results suggest that these phases meet  
only at zero temperature and, as expected, that the 
FD and AD transitions are continuous.

    Initially, let us comment on the Griffiths symmetry (see item
$(v)$ in the Introduction). For $J=0$ and $K<0$, we reobtain
the antiferromagnetic Ising model in the presence of a uniform
field. The phase diagram of this model is known \cite{dico}: it presents
an antiferromagnetic ordered phase and a paramagnetic phase. The
critical temperature between these two phases diminishes
as the field is increased, reaching zero for a critical field
$H_C=4 \mid J \mid$ (on the square lattice). The universality class for
finite field transitions is the same as for the zero-field Ising model, 
i.e. the flux along the boundary between the antiquadrupolar and
the disordered phases is towards the zero-field
fixed point. So, in an exact RG formalism one expects to obtain
a fixed point at \mbox{$(J=0,\; K=-1.763,\; \Delta=-2.832)$}, which
is stable along the phase boundary. Our results are depicted in Figure 4;
the exact result is shown as a dark circle while our evaluation of
the fixed point ($A^*$) is represented by a dark square.
Note that the shape of the  boundary is qualitatively correct,
while the location of the fixed point is different from the
expected result. Again this discrepancy is related to surface sites 
and one expects $A^*$ to approach the exact result
as the size of the cell grows. However, we obtain
the correct behavior in what regards universality class, i.e. the fixed
point is stable along the boundary (in fact, as we will comment
later, it is stable along the surface which separates the disordered
and antiquadrupolar phases; this surface is present for $J>0$ as
well).

   For $J < 0.4407$ the only ordered phase is the antiquadrupolar one,
while for $J \geq 0.4407$ a ferromagnetic ordered phase is present.
The attractor of the AD transition surface is the fixed point 
$(J^*,\; K^*,\; \Delta^*) =  (0,\; -3.23,\; -2.03)$; as discussed above,
this is not the exact result but it correctly represents the
antiquadrupolar phase, in the sense that  it belongs to the
$J=0$ plane; moreover, its eigenvalues show that the 
AD transition is a continuous one
(a first-order transition is indicated by $\lambda=b^d$,
where $\lambda$ is the eigenvalue associated with the field
conjugated to the order parameter, $b$ is the scaling parameter
and $d$ is the dimension of the system \cite{fisher}). This is the 
correct result, since this transition is in the same universality 
class of the zero-field Ising model. On the other hand, the 
FD boundary is attracted to the
$C$ fixed point (see Table I), which represents the spin-1/2
Ising model. Our procedure obtains this transition
as a continuous one, as well.

   In Figure 5 we show sections of constant $K/J < 0$. In $(a)$ the
behavior is representative of small values of $\mid K/J \mid$: 
the transition is always continuous, except for the zero-temperature
fixed point. Note that no reentrant behavior is obtained, contrarily to
the mean-field result (MF); our result is supported by the Monte Carlo
calculation of Reference \cite{japa} (not depicted in Figure $5a$).
In $(b)$ the $K/J = -2$ section is shown; as
discussed above, two ordered phases are present, namely the 
ferromagnetic and antiquadrupolar phases. The AD and FD transitions 
are both continuous
and their boundaries meet at zero temperature, where the transition is
first order. For smaller values of $K/J$ the antiquadrupolar phase
bulges out but the qualitative behavior is the same as for 
$K/J = -2$. 
Since for $K/J=-3$  and $\Delta/J=-2$ the antiferromagnetic
three-state Potts model on the square lattice is reobtained and this 
model does not order at finite temperatures, no phase transition
should be detected along this line: this is consistent with our 
results but not with the Monte Carlo result (MC).

   We also compare, in Figure $5b$, our results with Monte Carlo and mean-field calculations. The difference with respect to the latter
is expected but notice that the Monte Carlo calculation 
predicts a first-order transition between the two ordered
phases for $K/J=-2$.
This is in contradiction with our results; bearing
in mind the connection with the antiferromagnetic three-state
Potts model, the present approach yields the correct 
qualitative behavior (taking into account that the phase diagram
for $K/J=-3$ presents the same overall qualitative features 
as for $K/J=-2$). 

\section{Summary}
  
      We apply a  real-space RG procedure to study the BEG model. Our
results for $K>0$ are in excellent
qualitative and quantitative agreement with previous works.
For $K<0$, the phase diagram is qualitatively different from 
mean field results; since these correctly describe 
high-dimensional systems, while our work is on the
square lattice, this difference is expected.
Some discrepancy is found with Monte Carlo
results, in what concerns the existence of a first order transition
between antiquadrupolar and ferromagnetic phases; our results
do not predict this transition, which is consistent with the lack
of an ordered phase at finite temperatures for the 
antiferromagnetic three-state Potts model on the square lattice.

\section{Acknowledgments}

   We would like to thank Prof. D. Mukamel and Prof.
W. Figueiredo for helpful discussions and
the Departamento de F\'{\i}sica, Universidade Federal da
Para\'{\i}ba, for allowing the use of their computational
facilities at early stages of this work.

\newpage

    {\Large{\bf Table Caption}}

\vskip 1.5cm

      {\bf Table I:} Fixed points obtained in the present work 
(midle column) and exact results (right column: those with a subscript $p$
are not exact but refer to the RG calculation of \cite{ber1}).

\vskip 3cm

    {\Large{\bf Figure Captions}}

\vskip 1.5cm

   {\bf Figure 1:} Construction of a hierarchical lattice  adequate
to simulate the square lattice. Numbers 1 and 1' and 2 and 2'
denote terminal spins while 3, 4, 5, and 6 denote internal spins. 
($a$): original lattice, with parameters $J, K,$ 
and $\Delta$; ($b$) cell obtained from collapsing the terminal sites: it
is essential that collapsed sites 1 and 1' (as well as 
2 and 2') belong to the same sub-lattice; ($c$) renormalized lattice, 
with renormalized parameters $J', K',$ and $\Delta'$. 

   {\bf Figure 2:} Phase diagram for $K>0$. Wavy lines denote smooth
continuation of surfaces. Full lines denote  continuous transitions
and dotted lines denote first-order transitions (except the line
$PL$, which is a line of critical end points). $T_0 P$ is an ordinary
tricritical line and $GP$ is an isolated critical line.
$T$ is the attractor of tricritical transitions, $P$ is the
three-state ferromagnetic Potts model, and $G$ is the Griffiths
fixed-point (see text).

   {\bf Figure 3:} $J=0$ section of the phase diagram for
positive K, showing the
Ising fixed point (black circle); our evaluation of this point is 
indistinguishable
from the exact location (see Table 1). The broken line is the
exact result and the full curve represents the present approximation;
note that they are indistinguishable just below and above the $G$
fixed point.

   {\bf Figure 4:} $J=0$ portion of the phase-diagram for negative
$K$. $A^*$ is the fixed-point obtained in the present  
approximation, which should be compared to  the exact one
(dark circle). The flux along the boundary is towards $A^*$.

   {\bf Figure 5:} Phase diagram for constant and negative $K/J$.
Full (broken) lines represent continuous (first-order) transitions
and $f$, $d$ and $i$ stands
for the ferromagnetic, antiquadrupolar, and ferrimagnetic ordered phases
respectively. 
$(a): K/J=-1$: the behavior is representative of small values
of $K/J$. $(b): K/J=-2$: we compare our results ($RG$)
with mean-field ($MF$) and Monte Carlo ($MC$) calculations. In the latter,
filled (empty) circles denote continuous (first-order) phase
transitions. 

\newpage

\vskip 1.5cm

     {\Large{\bf Table I:}}

\vskip 1cm

\begin{tabular}{|c|c|c|} \hline
Fixed Point & Our results ($J^*,K^*,\Delta^*$) & Exact or previous result
\\ \hline 
   $C$  &  $(0.4407,-0.0831,-\infty)$  &   $(0.4407,K^*,-\infty)$    
\\ $G$  &  $(0,1.795,4.217)$           &   $(0,1.763,4.219)$       
\\ $L$  &  $(0.4407,\infty,2K^*+0.957)$&  $(0.4407,\infty,2K^*+1.078)_p$     
\\ $T$  &  $(1.470,0.0529,2.903)$      &   $(1.139,0.9944,4.245)_p$        
\\ $P$  &  $(0.5025,1.508,4.020)$      &   $(0.5025,1.508,4.020)$       
\\ $Fe$ &  $(\infty,\ln 2 - J^*,-\infty)$& $(\infty,K^*,-\infty)$      
\\ $F_2$&  $(0,\infty,2K^* + \ln 2))$  &   $(0,\infty,2K^* + \ln 2))$ 
\\ $Pa_+$ &$(0,0,-\infty)$             &   $(0,0,-\infty)$       
\\ $Pa_-$ &$(0,0,\infty)$              &   $(0,0,\infty)$            
\\ \hline
\end{tabular}

\end{document}